\documentclass[aps,twocolumn,showpacs]{revtex4}
\usepackage{graphicx}
\usepackage{subfigure}
\usepackage{bm}
\usepackage{amsmath}
\usepackage{amssymb}

\begin{document}
\title{Dynamics of Thermal Effects in the Spin-Wave Theory of Quantum Antiferromagnets}
\author{\'{A}ngel Rivas}
\email{anrivas@pas.ucm.es}
\altaffiliation[Fax: ]{+34 913945197}
\author{Miguel A. Martin-Delgado}
\affiliation{Departamento de F\'isica Te\'orica I, Facultad de Ciencias F\'{i}sicas, Universidad Complutense, 28040 Madrid, Spain}

\begin{abstract}
We derive a master equation that allows us to study non-equilibrium dynamics of a quantum antiferromagnet. By resorting to spin-wave theory, we obtain a closed analytic form for the magnon decay rates. These turn out to be closely related to form factors, which are experimentally accessible by means of neutron and Raman scattering. Furthermore, we compute the time evolution of the staggered magnetization showing that, for moderate temperatures, the magnetic order is not spoiled even if the coupling is fully isotropic.
\end{abstract}


\maketitle

\section{Introduction}
\label{sec:intro}

The properties of the quantum Heisenberg model play a fundamental role in the physics of many-body effects for models
defined by quantum Hamiltonians on a lattice, in several spatial dimensions \cite{interacting,strongly}. One of the first
non-perturbative methods devised to study the quantum Heisenberg model is known as spin-wave theory (SWT).  This is a type
of mean-field theory method that is especially suited to study the quantum fluctuations of interacting spins. The basic assumption
is the existence of a ground state that spontaneously breaks the global symmetry of the Heisenberg Hamiltonian. In this case, it
corresponds to rotational symmetry SO(3) about an arbitrary axis. In SWT this symmetry is broken by fixing a preferred axis called
magnetization axis of the ground state, and excitations appear in the form of fluctuations from the fixed direction. These are the
Goldstone bosons of this spontaneously breaking mechanism and represent the magnon modes propagating as spin waves in the quantum system.
However, spatial dimensionality is crucial in order to have a well-defined semiclassical expansion in the parameter $1/S$, where $S$ is the
total spin at each site of the system lattice. Namely, in a quantum antiferromagnet the spatial dimension of the lattice has to be large enough in order to sustain the assumption of a given order in the ground state. Otherwise, strong quantum fluctuations in one-dimensional lattices break the long-range order and makes the SWT invalid. However, many interesting systems are materials in 3D, and SWT provides very good approximations to their observable quantities.

SWT has been extensively developed in many aspects. It has become by now a standard and reference tool in order to have
a good approximate description of quantum antiferromagnetic systems, whenever the validity of its application is justified.

To the best of our knowledge, there is an important aspect of SWT that remains vaguely explored, namely, the modification of SWT
in order to adapt it to describe the natural interaction of a quantum antiferromagnet with an external or surrounding thermal bath
that is interacting with it. A typical example is provided by the phonons of the lattice, where the quantum spins are located.
This is a basic and fundamental problem since it entails the description of both dynamical effects, i.e. time-dependent, as well as finite-temperature effects outside the state of thermal equilibrium.

Embedding thermal fluctuations in the dynamics of a system may be approached from several points of view. For instance, in the classical domain, it is common to consider the effect of a noisy magnetic thermal field acting on the Heisenberg Hamiltonian \cite{Brown}. However, that situation is different from what we focus in this work, where the noise is described from a microscopic model based on thermal excitation of the surrounding environment. The branch of the quantum theory that deals with this kind of problems is the theory of open quantum systems \cite{libro,Weiss,BrPe,GardinerZoller} that plays a fundamental role in quantum information theory
\cite{NC,rmp}. From this point of view, the quantum magnet is considered as an open system, which exchanges energy with its environment.

The best method to describe an open system strongly depends on the explicit nature of each situation. For example, recently an approach based on the non-equilibrium functional renormalization group has been proposed for the study of the thermalization of a magnon gas in contact with a thermal phonon bath \cite{Hick}. In this work, we have applied the Davies formalism, which is a suitable description of an open system weakly interacting with a large environment. One of its main features is that it allows us to derive an evolution equation for any spin observable of the quantum antiferromagnet coupled to a generic thermal bath at a certain temperature $T$. Namely, it provides us with an equation for the evolution of the density matrix $\rho(t)$. Furthermore, as a consequence of how this fundamental equation is obtained, a series of  interesting results for the enlarged SWT have been obtained: i/ the quantum antiferromagnet thermalizes towards the Gibbs state for long enough times; ii/ the decay rate of this thermalization process can be obtained in a closed analytical form as a function of the lattice momentum; iii/ the thermal bath cannot be arbitrary in order to ensure the convergence of any observable to its thermal value, but it has to belong to the class of super-ohmic baths with specific parameters, depending on the quantum antiferromagnet; iv/ the staggered magnetization can be computed analytically and we can obtain its behaviour with time and temperature, thereby unveiling the fate of the antiferromagnetic order parameter; and v/ the thermal evolution of the magnon form factor can also be computed explicitly. These quantities are of physical importance and observable in inelastic neutron scattering \cite{Shirane87,Aeppli89,Perring96} and Raman experiments \cite{Lyons88} for instance.

Let us emphasize that the framework of our investigations is the out-of-equilibrium dynamics in a spin-wave system coupled to a bosonic thermal bath. The methodology employed is the master equation formalism for open quantum systems. With this combination of dynamics and methodology, we have found new behavior for the spin-wave decay rates at finite temperature, that have not been treated previously. Earlier studies of damping effects in spin waves at finite temperature, such as \cite{Stinchcombre}, rely on the use of the Gibbs state at different temperatures. Nevertheless, let us note that, in our study, the magnons are damped while the system is approaching the Gibbs state, not once the system is in equilibrium with the environment at some temperature. It is this type of new physics that we can address in a different way than the previous investigations.

As for the physical nature of the coupling between the system spin waves and the bosonic external bath, we may consider at least two possible practical realizations.

a/ Quantum simulations with optical lattices: the experimental realization of a controlled Mott insulator to superfluid transition with cold atoms in an optical trap \cite{Simulator} has opened the field to quantum simulations of new physics in a range of parameters and types of couplings that are not easy to find in nature, but they are feasible to engineer.

b/ Interaction with phonons in a crystal lattice: although it is natural to think of lattice phonons in a condensed matter system as a candidate for the bosonic coupling to the spin waves, this possibility comes with several caveats. First, we have employed spin-wave theory in the first-order approximation (linear spin-wave theory (LSWT)). While there are theoretical studies that support the use of these approximations in 3D quantum spin systems at finite temperature \cite{Oguchi}, not all materials of this class exhibit a behavior according to LSWT. Nevertheless, it is possible to find several types of compounds whose magnons behave very well as predicted by LSWT \cite{a,b,c,d}, and these are candidates for the application of our results. However, we also point out that in addition to the magnon-boson channel studied in this paper, real materials may also have other decay channels due to magnon-electron interactions or coupled orbital-lattice fluctuations \cite{canales} that are outside our current framework.

This paper is organized as follows: in Sect.\ref{sec:SWT}, we review the linear spin-wave theory and establish our notation.
In Sect.\ref{sec:bosonic_environment}, we describe the microscopic coupling of the SWT Hamiltonian with a Hamiltonian bath of bosonic operators.
In Sect.\ref{sec:master_equation}, we derive the complete master equation for describing the evolution and thermal effects of the system due to its interaction with the bath.
In Sect.\ref{sec:observables}, we compute relevant observables under the above conditions, such as the staggered magnetization, two-spin correlators, and form factors.
Sect.\ref{sec:conclusions} is devoted to conclusions.
We refer to appendix \ref{appendixA} for expressions of the time evolution of the first and second moments and to appendix \ref{appendixB} for the detailed calculation involving the two-spin spatial correlation functions.

\section{Spin Wave Theory for Quantum Antiferromagnets}
\label{sec:SWT}

\begin{figure}[t!]
\begin{center}
\includegraphics[width=0.45\textwidth]{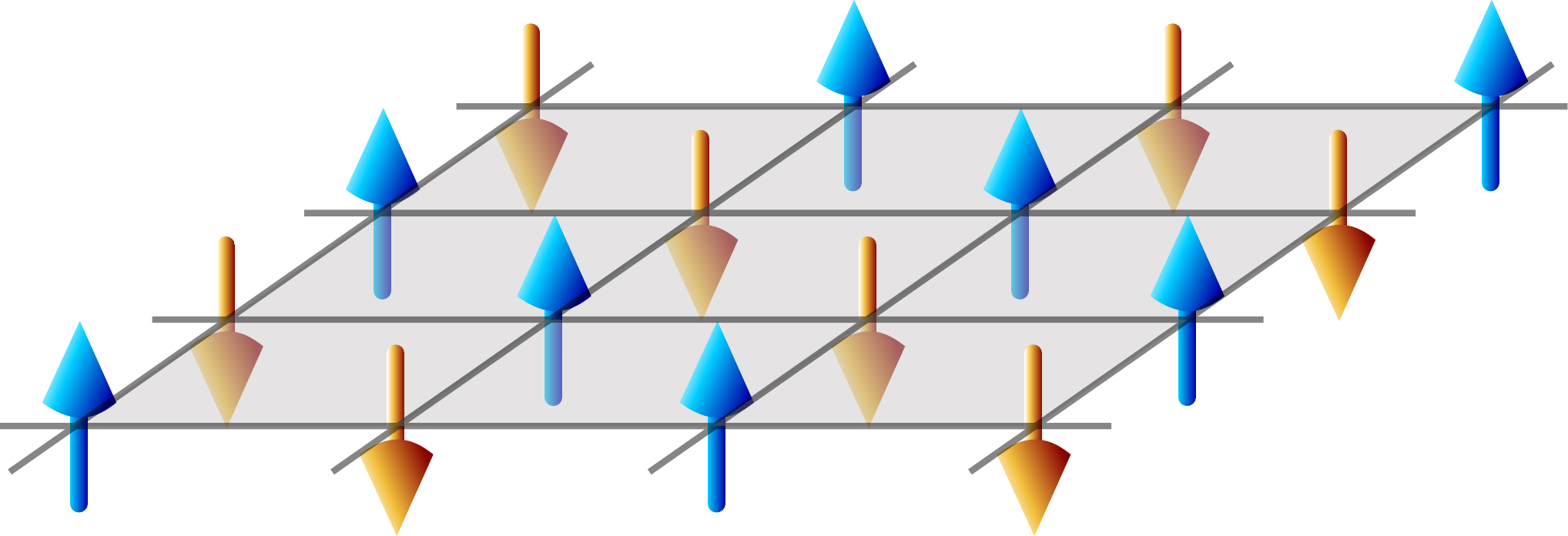}
\end{center}
\caption{Arrangement of a quantum antiferromagnet in the N\'eel state on a square lattice in 2D. The color of the spins denotes the two different sublattices, $A$ (blue arrows) and $B$ (orange arrows). The true ground state is close to this staggered configuration; however, there is a slight disarrangement in the orientation of the spins due to quantum fluctuations.}\label{fig1}
\end{figure}

First, let us briefly recall the spin-wave theory for quantum antiferromagnets and for establishing our notation. The system consists of a lattice with a spin $S$ on every vertex. The Hamiltonian contains only two-body terms between the first neighbors according to the Heisenberg interaction
\begin{equation}\label{Hantiferro}
H_S=J\sum_{\langle \bm{r},\bm{r}'\rangle}\bm{S}_{\bm{r}}\cdot\bm{S}_{\bm{r}'},
\end{equation}
with $J>0$ for antiferromagnetism. The phenomenology displayed by this Hamiltonian strongly depends on the morphology of the lattice. Particularly, if the lattice is bipartite (i.e. we can define two sublattices $A$ and $B$ in such a way that the first neighbors of a $A$ belong to $B$ and vice versa, see figure \ref{fig1} for an illustration of the two-dimensional case), the ground state is close to a staggered spin configuration known as N\'eel state. However, if the lattice is not bipartite (e.g. triangular lattice), the system becomes frustrated, and no simple configuration is found to be a ground state for the diagonal part of the Hamiltonian \eqref{Hantiferro}. For our purposes, we shall consider a 3D square lattice.

The diagonalization of the Hamiltonian (\ref{Hantiferro}) is not an easy task, and no exact solutions are known for spatial dimensions $d\geq 2$ or for spins $S\geq 1$ in $d=1$.  Thus,  approximation methods become very useful. Probably the most fundamental of them is based on the Holstein--Primakoff approximation \cite{Review,HolsteinPrimakoff40} and leads to the so-called spin-wave theory \cite{Anderson52}, which is also applicable to ferromagnets \cite{Kubo88}. This method rewrites the spin operators in terms of bosonic annihilation and creation operators, $a$ and $a^\dagger$, $[a,a^\dagger]=1$. Concretely, for $\bm{r}$ in the sublattice $A$
\begin{eqnarray} \label{latticeA}
S_{\bm{r}}^+&=&\sqrt{2S}f_S(a^\dagger_{\bm{r}}a_{\bm{r}})a_{\bm{r}},\nonumber\\
S_{\bm{r}}^-&=&\sqrt{2S}a^\dagger_{\bm{r}}f_S(a^\dagger_{\bm{r}}a_{\bm{r}}),\\
S_{\bm{r}}^z&=&S-a^\dagger_{\bm{r}}a_{\bm{r}},\nonumber
\end{eqnarray}
and for $\bm{r}$ in the sublattice $B$
\begin{eqnarray}\label{latticeB}
S_{\bm{r}}^+&=&\sqrt{2S}b^\dagger_{\bm{r}}f_S(b^\dagger_{\bm{r}}b_{\bm{r}}),\nonumber\\
S_{\bm{r}}^-&=&\sqrt{2S}f_S(b^\dagger_{\bm{r}}b_{\bm{r}})b_{\bm{r}},\\
S_{\bm{r}}^z&=&b^\dagger_{\bm{r}}b_{\bm{r}}-S,\nonumber
\end{eqnarray}
with
\begin{equation}
f_S(x)=\left(1-\frac{x}{2S}\right)^{1/2}.
\end{equation}
By writing the Hamiltonian (\ref{Hantiferro}) at the first order in $f_S(x)$, we obtain the so-called linear spin-wave theory:
\begin{eqnarray}\label{HLSW}
H_{LSW}=&J\left[-NdS^2+2dS\sum_{\bm{r}}(a^\dagger_{\bm{r}}a_{\bm{r}}+b^\dagger_{\bm{r}}b_{\bm{r}})\right.\nonumber\\
&+\left.S\sum_{\langle \bm{r},\bm{r}'\rangle}(a_{\bm{r}}b_{\bm{r}'}+a^\dagger_{\bm{r}}b^\dagger_{\bm{r}'})\right].
\end{eqnarray}
This approximation is valid to describe states, where $\langle f_S(a^\dagger_{\bm{r}}a_{\bm{r}})\rangle=\langle f_S(b^\dagger_{\bm{r}}b_{\bm{r}})\rangle\simeq1$, and thus, they also verify
\begin{equation}\label{validitySW}
\langle a^\dagger_{\bm{r}}a_{\bm{r}}\rangle,\ \langle b^\dagger_{\bm{r}}b_{\bm{r}}\rangle\ll 2S.
\end{equation}
This is the self-consistent condition characteristic of this mean-field theory method.

The Hamiltonian $H_{LSW}$ is quadratic in boson operators, so in order to diagonalize it, we take Fourier transform:
\begin{eqnarray}\label{fourierab}
a_{\bm{r}}&=&\sqrt{\frac{1}{N_A}}\sum_{\bm{k}}{\rm e}^{-{\rm i} \bm{k}\cdot\bm{r}}a_{\bm{k}},\\
b_{\bm{r}}&=&\sqrt{\frac{1}{N_B}}\sum_{\bm{k}}{\rm e}^{{\rm i} \bm{k}\cdot\bm{r}}b_{\bm{k}},
\end{eqnarray}
with $N_A=N_B=N/2$ for a square lattice and the lattice wave vector takes on the following discretized values:
\begin{equation}
\bm{k}=\frac{2\pi \bm{m}}{N_{A,B}}=\frac{4\pi \bm{m}}{N},\ \bm{m}\in A,B.
\end{equation}
Then, the first term of $H_{LSW}$ is easy to compute, given the orthonormalization rule $\frac{2}{N}\sum_{\bm{r}\in A,B}{\rm e}^{{\rm i}\bm{k}\cdot\bm{r}}=\delta_{\bm{k},0}$. For the second one, we parameterize $\bm{r}'$ neighbor to $\bm{r}$ as $\bm{r}'=\bm{r}+\hat{\bm{r}}_\mu$, where $\hat{\bm{r}}_\mu$ is the unit vector in the $\mu$ direction, which in $d=3$ and starting from the first site, can be $(1,0,0)$, $(0,1,0)$ or $(0,0,1)$. Thus, we obtain
\begin{eqnarray}\label{HLSWk}
H_{LSW}=&J\left[-NdS^2+2S\sum_{\bm{k}}d(a^\dagger_{\bm{k}}a_{\bm{k}}+b^\dagger_{\bm{k}}b_{\bm{k}})\right.\nonumber\\
&+\left.\xi_{\bm{k}}(a_{\bm{k}}b_{\bm{k}}+a^\dagger_{\bm{k}}b^\dagger_{\bm{k}})\right],
\end{eqnarray}
where
\[
\xi_{\bm{k}}=\sum_\mu \cos(\bm{k}\cdot\hat{\bm{r}}_\mu).
\]

Next step is to perform a Bogoliubov transformation to new boson operators $\alpha_{\bm{k}}$ and $\beta_{\bm{k}}$:
\begin{eqnarray}
a_{\bm{k}}&=&\cosh(\theta_{\bm{k}})\alpha_{\bm{k}}-\sinh(\theta_{\bm{k}})\beta^\dagger_{\bm{k}},\label{Bologiubova}\\
b_{\bm{k}}&=&-\sinh(\theta_{\bm{k}})\alpha^\dagger_{\bm{k}}+\cosh(\theta_{\bm{k}})\beta_{\bm{k}}.\label{Bologiubovb}
\end{eqnarray}
The function $\theta_{\bm{k}}$ is chosen so that the coefficient of $\alpha_{\bm{k}}\beta_{\bm{k}}$ and $\alpha^\dagger_{\bm{k}}\beta^\dagger_{\bm{k}}$ is zero:
\begin{equation}\label{xi}
\tanh(2\theta_{\bm{k}})=\frac{\xi_{\bm{k}}}{d}.
\end{equation}
With this choice, the Hamiltonian of the system is diagonalized:
\begin{equation}\label{HLSWkdiag}
H_{LSW}=E_0^0+\sum_{\bm{k}}\omega(\bm{k})(\alpha^\dagger_{\bm{k}}\alpha_{\bm{k}}+\beta^\dagger_{\bm{k}}\beta_{\bm{k}}).
\end{equation}
Here, the energy dispersion relation is
\begin{equation}\label{wk}
\omega(\bm{k})=2JS\sqrt{d^2-\xi_{\bm{k}}^2},
\end{equation}
and $E_0^0$ is a constant
\[
E_0^0=-JNS\left[dS+\frac{2}{N}\sum_{\bm{k}}\left(d-\sqrt{d^2-\xi_{\bm{k}}^2}\right)\right].
\]

In summary, we have transformed the intricate Hamiltonian (\ref{Hantiferro})  with interaction terms into another approximate Hamiltonian, which is just a collection of uncoupled harmonic oscillators, and hence, it is easy to write the whole spectrum analytically. The excitations of these harmonic oscillators are called ``magnons'', because they represent the minimal collective magnetic excitation of the spin lattice.

\section{Interaction with a bosonic environment}
\label{sec:bosonic_environment}

The antiferromagnetic system may be affected by a dissipative dynamics due to the interaction with its environment. In principle, the most common source of dissipation will be bosonic excitations in the lattice (e.g. phonons). Thus, the interaction Hamiltonian will be given typically by the so-called spin-boson model \cite{Weiss,Leggett}, $V\propto \bm{S}\cdot \bm{R}$, where $\bm{S}=(S^x,S^y,S^z)$ is the spin vector and $\bm{R}=(X,Y,Z)$ is the position operators of the bosonic environment. Other types of coupling (e.g. \cite{Pincus}) could eventually be taken into account. In addition, and as a first proposal, we assume a local environmental model:
\begin{eqnarray}\label{V}
V&=&\sum_j\sum_{\bm{r}} g(\omega_j)\left[S^x_{\bm{r}}(A_{\bm{r},j}^x+A^{x\dagger}_{\bm{r},j})\right.\nonumber\\
&+&\left.S^y_{\bm{r}}(A_{\bm{r},j}^y+A^{y\dagger}_{\bm{r},j})+S^z_{\bm{r}}(A_{\bm{r},j}^z+A^{z\dagger}_{\bm{r},j})\right].
\end{eqnarray}
Here, $A$ and $A^\dagger$ stand for annihilation and creation operators of the environmental boson modes, and we have assumed that the coupling function $g(\omega_j)$ is isotropic and the same for every member of the lattice. On the other hand, the Hamiltonian of the environment is
\begin{equation}
H_E=\sum_j \sum_{\bm{r}} \omega_j(A^{x\dagger}_{\bm{r},j}A_{\bm{r},j}^x+A^{y\dagger}_{\bm{r},j}A_{\bm{r},j}^y+A^{z\dagger}_{\bm{r},j}A_{\bm{r},j}^z),
\end{equation}
which is written as
\begin{equation}
H_E=\sum_j \sum_{\bm{k}} \omega_j(A^{x\dagger}_{\bm{k},j}A_{\bm{k},j}^x+A^{y\dagger}_{\bm{k},j}A_{\bm{k},j}^y+A^{z\dagger}_{\bm{k},j}A_{\bm{k},j}^z),
\end{equation}
after taking Fourier transform.

In linear spin-wave theory approximation, the interaction term reads
\begin{eqnarray*}
V_{LSW}&=\sum_j g(\omega_j)\left\{\sum_{\bm{r}\in A} \left[\sqrt{\frac{S}{2}}(a_{\bm{r}}+a^\dagger_{\bm{r}})(A_{\bm{r},j}^x+A^{x\dagger}_{\bm{r},j})\right.\right.\\
&-{\rm i}\sqrt{\frac{S}{2}}(a_{\bm{r}}-a^\dagger_{\bm{r}})(A_{\bm{r},j}^y+A^{y\dagger}_{\bm{r},j})\\
&+\left. (S-a^\dagger_{\bm{r}}a_{\bm{r}})(A_{\bm{r},j}^z+A^{z\dagger}_{\bm{r},j})\right]\\
&+\sum_{\bm{r}\in b} \left[\sqrt{\frac{S}{2}}(b_{\bm{r}}+b^\dagger_{\bm{r}})(A_{\bm{r},j}^x+A^{x\dagger}_{\bm{r},j})\right.\\
&+{\rm i}\sqrt{\frac{S}{2}}(b_{\bm{r}}-b^\dagger_{\bm{r}})(A_{\bm{r},j}^y+A^{y\dagger}_{\bm{r},j})\\
&+\left.\left.(b^\dagger_{\bm{r}}b_{\bm{r}}-S)(A_{\bm{r},j}^z+A^{z\dagger}_{\bm{r},j})\right]\right\}.
\end{eqnarray*}
Now, the whole Hamiltonian has become much more involved than the original spin-wave theory Hamiltonian.
However, we can consider a simplified version of the interaction Hamiltonian $V_{LSW}$ based on the following two facts:
\begin{itemize}
\item The terms $a^\dagger_{\bm{r}}a_{\bm{r}}$ and $b^\dagger_{\bm{r}}b_{\bm{r}}$ are negligible in comparison to the others in the regime where the spin-wave theory is valid (\ref{validitySW}).
\item We ignore the term $S(A_{\bm{r},j}+A^{\dagger}_{\bm{r},j})$ because it is a fast oscillator, which we may neglect in the weak coupling limit, see below.
\end{itemize}
Therefore, after taking Fourier transform, we arrive at
\begin{eqnarray}\label{VLSWk}
V_{LSW}&=&\sqrt{\frac{S}{2}}\sum_j\sum_{\bm{k}} g(\omega_j)\left[(a_{\bm{k}}+b^\dagger_{\bm{k}})\right.\\
&\times & \left.(A_{-\bm{k},j}^x+A^{x\dagger}_{\bm{k},j}-{\rm i}A_{-\bm{k},j}^y-{\rm i}A^{y\dagger}_{\bm{k},j})+{\rm h.c.}\right].\nonumber
\end{eqnarray}
Finally, the Bogoliubov  transformation of Eqs. (\ref{Bologiubova}) and (\ref{Bologiubovb}) leads to
\begin{eqnarray}\label{VLSWkfinal}
V_{LSW}&=&\sqrt{\frac{S}{2}}\sum_j\sum_{\bm{k}} g(\omega_j)\left(\frac{d-\xi_{\bm{k}}}{d+\xi_{\bm{k}}}\right)^{1/4}\nonumber\\
&\times&\left[(\alpha_{\bm{k}}+\beta^\dagger_{\bm{k}})(A_{-\bm{k},j}^x+A^{x\dagger}_{\bm{k},j}-{\rm i}A_{-\bm{k},j}^y-{\rm i}A^{y\dagger}_{\bm{k},j})\nonumber\right.\\
&+&\left.{\rm h.c.}\right].
\end{eqnarray}

\section{Master equation for a thermal environment}
\label{sec:master_equation}

The dynamics of the system and the environment is given by the von Neumann equation
\begin{equation}
\frac{d\rho}{dt}=-\frac{\text{i}}{\hbar}[H,\rho],
\end{equation}
where
\begin{equation}
H=H_{LSW}+H_E+V_{LSW}.
\end{equation}
We aim at writing a dynamical equation for the state of the system $\rho_S=\mathrm{Tr}_E(\rho)$, where the trace is taken over the environment degrees of freedom. This task is generally quite complicated. However, we are particularly interested in describing how the system evolves to the Gibbs state because of the lack of insulation, and such a case is expected to happen for a large environment in thermal equilibrium (a ``bath'') with a small coupling constant. Under these conditions, an equation, called the master equation, can be found by resorting to perturbation theory \cite{Davies}.

The initial state of the environment is then written as
\begin{eqnarray}
\rho_E&=&Z^{-1}{\rm e}^{-\beta H_E}\\
&=&Z^{-1}{\rm e}^{-\beta\sum_j \sum_{\bm{k}} \omega_j(A^{x\dagger}_{\bm{k},j}A_{\bm{k},j}^x+A^{y\dagger}_{\bm{k},j}A_{\bm{k},j}^y+A^{z\dagger}_{\bm{k},j}A_{\bm{k},j}^z)}\nonumber,
\end{eqnarray}
where $Z=\mathrm{Tr}\left({\rm e}^{-\beta H_E}\right)$ is the partition function with $\beta=1/k_{\rm B}T$. From now on, we shall use natural units $\hbar=k_\mathrm{B}=1$.

Due to the Riemann–-Lebesgue lemma \cite{libro}, for small enough \cite{RWA} coupling $g(\omega_j)$, we can safely neglect the counter-rotating terms in (\ref{VLSWkfinal}):
\begin{eqnarray}
V_{LSW}=&\sqrt{S}\sum_j\sum_{\bm{k}} g(\omega_j)\left(\frac{d-\xi_{\bm{k}}}{d+\xi_{\bm{k}}}\right)^{1/4}\left[\alpha_{\bm{k}}(A^{x\dagger}_{\bm{k},j}-{\rm i}A^{y\dagger}_{\bm{k},j})\right.\nonumber\\
&+\left.\beta^\dagger_{\bm{k}}(A_{-\bm{k},j}^x-{\rm i}A^{y}_{-\bm{k},j})+{\rm h.c.}\right].
\end{eqnarray}
Now, the problem becomes equivalent to two collections of uncoupled harmonic oscillators given by their operators $\alpha_{\bm{k}}$ and $\beta_{\bm{k}}$, which are coupled to a set of independent environments characterized by $\bm{k}$. The standard tools to obtain a master equation for a weak interaction with the environment can be found in references \cite{libro,Weiss,BrPe,GardinerZoller}. If we apply those techniques to this system, we arrive at
\begin{widetext}
\begin{eqnarray}\label{Master}
\frac{d\rho}{dt}=\mathcal{L}(\rho)=&-&{\rm i}[H_{LSW},\rho]
+\sum_{\bm{k}} \gamma_{\bm{k}}(\bar{n}_{\bm{k}}+1)\left(\alpha_{\bm{k}}\rho\alpha^\dagger_{\bm{k}}-\frac{1}{2}\{\alpha^\dagger_{\bm{k}}\alpha_{\bm{k}},\rho\}+\beta_{\bm{k}}\rho\beta^\dagger_{\bm{k}}-\frac{1}{2}\{\beta^\dagger_{\bm{k}}\beta_{\bm{k}},\rho\}\right)\nonumber\\
&+&\gamma_{\bm{k}}\bar{n}_{\bm{k}}\left(\alpha^\dagger_{\bm{k}}\rho\alpha_{\bm{k}}-\frac{1}{2}\{\alpha_{\bm{k}}\alpha^\dagger_{\bm{k}},\rho\}+\beta^\dagger_{\bm{k}}\rho\beta_{\bm{k}}-\frac{1}{2}\{\beta_{\bm{k}}\beta^\dagger_{\bm{k}},\rho\}\right)
\end{eqnarray}
\end{widetext}
Here
\begin{equation}\label{decayrates}
\gamma_{\bm{k}}:=2\pi S\sqrt{\frac{d-\xi_{\bm{k}}}{d+\xi_{\bm{k}}}} \mathcal{J}(\omega(\bm{k})),
\end{equation}
where $\mathcal{J}(\omega)=\sum_jg^2(\omega)\delta(\omega-\omega_j)$ is the so-called spectral density of the bath. This one, for solid-state environments, is usually parameterized in the continuous limit \cite{Weiss,Leggett} as
\begin{equation}
\mathcal{J}(\omega)=\alpha\omega^s\omega_c^{s-1}{\rm e}^{-\omega/\omega_c},
\label{spectral_density}
\end{equation}
where $\alpha$ accounts for the strength of the coupling and $\omega_c$ is the cut-off frequency of the bath. Typically three cases are distinguished: $s>1$ (super-ohmic), $s=1$ (ohmic), and $s<1$ (sub-ohmic). The other quantity $\bar{n}_{\bm{k}}$ is the mean number of phonons in the bath with frequency $\omega({\bm{k}})$:
\begin{equation}
\bar{n}_{\bm{k}}:=[\exp(\omega({\bm{k}})/T)-1]^{-1}.
\end{equation}

\subsection{Approach to the Equilibrium}

By construction \cite{Davies}, the Gibbs state $\rho_{\mathrm{th}}=Z^{-1}{\rm e}^{-H_{LSW}/T}$, at the same temperature $T$ as the bath, is the steady state of equation (\ref{Master}), i.e. $\mathcal{L}(\rho_{\mathrm{th}})=0$. This is straightforwardly verified by taking into account that
\begin{eqnarray}
{\rm e}^{-H_{LSW}/T}\alpha_{\bm{k}}&=&{\rm e}^{\omega(\bm{k})/T}\alpha_{\bm{k}}{\rm e}^{-H_{LSW}/T},\\
{\rm e}^{-H_{LSW}/T}\beta_{\bm{k}}&=&{\rm e}^{\omega(\bm{k})/T}\beta_{\bm{k}}{\rm e}^{-H_{LSW}/T}.
\end{eqnarray}
Moreover, any initial state of the system becomes closer and closer to this Gibbs state during time evolution.

We have thus constructed a dynamical equation to describe the thermal relaxation process of a quantum antiferromagnet. Remember that for the spin-wave theory to make sense, the number of magnons has to be small (\ref{validitySW}), so for large bath temperatures this treatment is not valid in the long-time limit where the system approaches the Gibbs state (which contains a large number of magnons for large $T$). However the predictions of equation (\ref{Master}) should also agree reasonably well with the exact ones at short times.

\subsection{Magnon Decay Rates}
\begin{figure}[t]
\begin{center}
\includegraphics[width=0.5\textwidth]{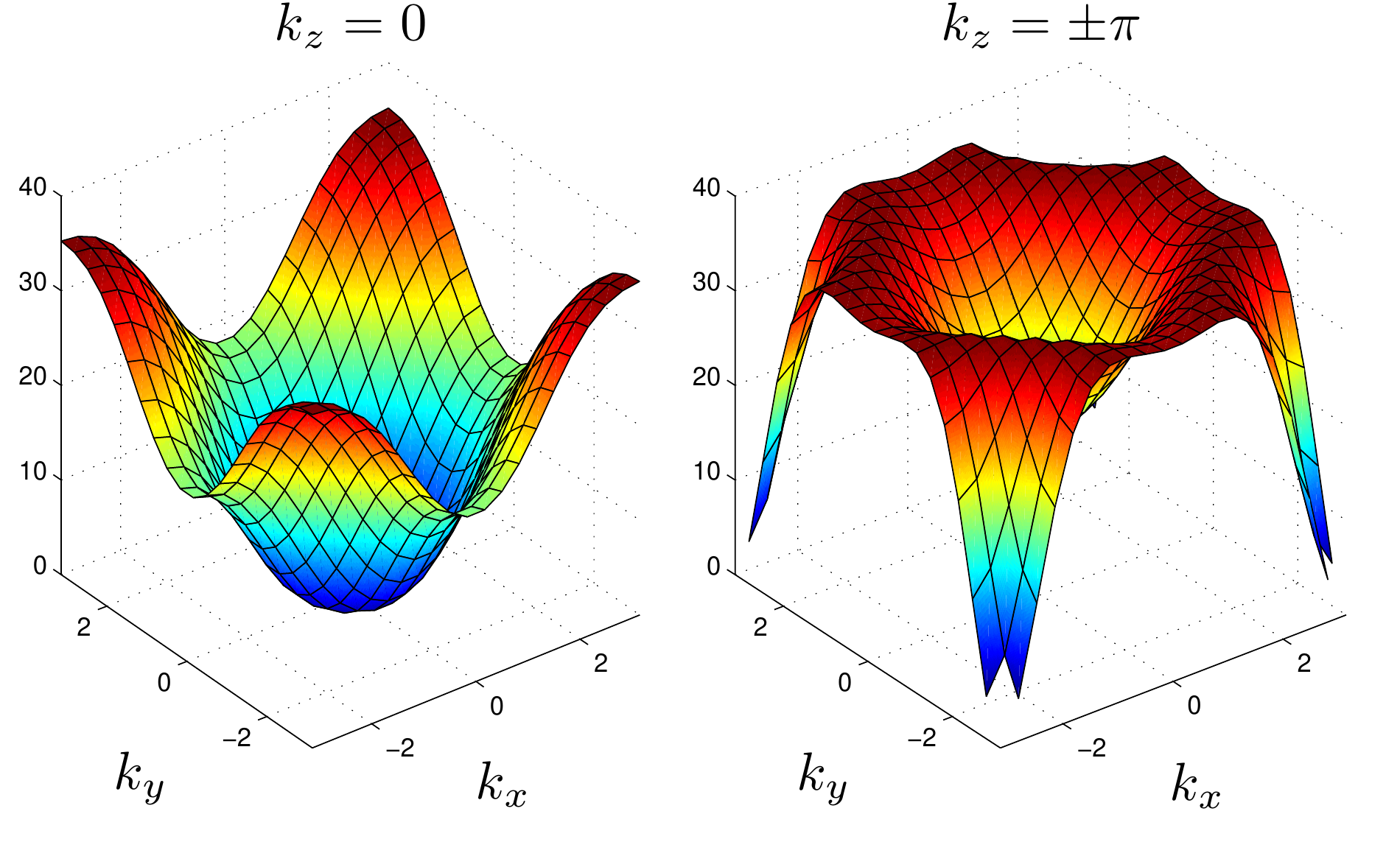}
\end{center}
\caption{Magnon decay rate (\ref{decayrates}) in the first Brillouin zone. The surface for $k_z=0$ and $k_z=\pm\pi$ is depicted on the left and right, respectively.}\label{fig2}
\end{figure}

A remarkable property of this system is that every exponent is not allowed in the spectral density \eqref{spectral_density} in order to obtain finite results for many-body observables. This is because quantities, such as magnon decay rates (\ref{decayrates}) and the thermal number of phonons, become infinite for certain values of $\bm{k}$, so for those values, the spectral density has to approach zero fast enough. Particularly, it requires a super-ohmic spectral density. It is worth recalling here that this kind of problems may also arise in simpler systems, for instance, in a single spin when subject to a pure dephasing environment (see \cite{libro}). The concrete values of the rest of parameters of $\mathcal{J}(\omega)$ are not very relevant for our purposes as we always assume to be in a sufficiently weak interaction regime \cite{Rivas09}; we shall take
\begin{equation}
s=3,  \quad  \alpha=J/10, \quad \omega_c=\max_{\bm{k}}\omega(\bm{k})=2JSd.
\label{spectral_choice}
\end{equation}

In figure \ref{fig2}, we have represented two sheets of the magnon decay rate (\ref{decayrates}) in the first Brillouin zone. On the one hand, we note that the magnon decay rate vanishes on the origin and on the eight corners of the Brillouin zone $\bm{k}=(\pm\pi,\pm\pi,\pm\pi)$. Therefore, magnons with these momenta are not affected by the presence of the thermal bath. From a quantum information point of view, the subspace
\begin{equation}
\mathcal{S}={\rm span}\{|n_{\bm{k}}\rangle|\bm{k}=(\pm\pi,\pm\pi,\pm\pi)\}
\end{equation}
is a decoherence free subspace, where we can store information robustly. Note that this is true independently of the temperature $T$ and the number of spins $N$.

On the other hand, the magnon decay rate reaches the maximum value on the points of a sphere of radius $r\simeq0.947$ centered just at these eight minimum points $\bm{k}=(\pm\pi,\pm\pi,\pm\pi)$. Between both cases, there is a transition that we have tried to illustrate by taking the values $k_z=0,\pm\pi$ in the figure (given the symmetry of the decay rate, we can use $k_{x,y}$ instead of $k_z$, leading to the same figures).

From (\ref{Master}), it is possible to compute the evolution of any combination of $\alpha_{\bm{k}}$ and $\beta_{\bm{k}}$ in the Heisenberg picture. We give the result for the evolution of the first and second moments in Appendix \ref{appendixA}.  Those expressions allow us to compute the evolution of any spin operator in the quantum antiferromagnet and relevant observables constructed out of them.

\section{Dynamics of relevant observables}
\label{sec:observables}

In this section, we study the time evolution of some properties that have special interest in the description of a quantum antiferromagnet. For concreteness, we have selected two of them: the staggered magnetization and the spin correlation functions.

\subsection{Staggered magnetization}
Due to the isotropy of Hamiltonian (\ref{Hantiferro}), one may expect the ground state also to be symmetric under rotations. However, as we have already mentioned, the ground state turns out to be close to the N\'eel state, which has clearly a privileged orientation. This is an example of spontaneous symmetry breaking \cite{Review} . The figure of merit to compute this order in a quantum antiferromagnet is the expectation value of the staggered magnetization operator:
\begin{equation}
\hat{m}^\mathrm{st}_z=\frac{1}{N}\sum_{\bm{r}}(-1)^{\Vert\bm{r}\Vert}S^z_{\bm{r}},
\end{equation}
which in the thermodynamic limit reads
\begin{equation}
m^\mathrm{st}=\lim_{N\rightarrow\infty}\langle\hat{m}^\dagger_z\rangle=\lim_{N\rightarrow\infty}\frac{1}{N}\sum_{\bm{r}}(-1)^{\Vert\bm{r}\Vert}\langle S^z_{\bm{r}}\rangle.
\end{equation}
By using the equations (\ref{latticeA}) and (\ref{latticeB}), we may write this operator as
\begin{eqnarray}
\hat{m}^\mathrm{st}_z&=&\frac{1}{N}\sum_{\bm{r}}(S-n_{\bm{r}})\nonumber\\
&=&S-\frac{1}{N}\sum_{\bm{r}}\hat{n}_{\bm{r}}=S-\frac{1}{N}\sum_{\bm{k}}\hat{n}_{\bm{k}},
\end{eqnarray}
with $\hat{n}_{\bm{k}}=\hat{n}_{\bm{k}}^{(a)}+\hat{n}_{\bm{k}}^{(b)}$. Note that $\bm{k}=2\pi\bm{m}/(N/2)$, where $\bm{m}$  varies two by two instead of one by one. So in the thermodynamic limit
\begin{equation}
\lim_{N\rightarrow\infty}\frac{1}{N}\sum_{\bm{k}}=\frac{1}{2(2\pi)^3}\int_\mathrm{B.Z.} d\bm{k}.
\end{equation}
Here, B.Z. stands for  the first Brillouin zone, and the extra factor $1/2$ appears because of the double spacing between consecutive $\bm{k}$ on the left-hand side. Thus, the staggered magnetization becomes
\begin{equation}\label{stm}
m^\mathrm{st}=S-\frac{1}{16\pi^3}\int_\mathrm{B.Z.} d\bm{k}\langle\hat{n}_{\bm{k}}\rangle.
\end{equation}

When the system is interacting with a thermal bath, the staggered magnetization approaches in time to its thermal value. This is exactly zero for any $T\neq0$ due to the Mermin-Wagner theorem \cite{MerminWagner} in 1D and 2D; however, that is not the case in 3D. Additionally, note that the interaction Hamiltonian (\ref{V}) is also isotropic, so it is not trivial to find also magnetic order when the quantum antiferromagnet is not isolated.

From the master equation (\ref{Master}), we are able to visualize how staggered magnetization varies as a function of time. For this aim, we just need to find the evolution of the observables $\hat{n}_{\bm{k}}$. In terms of the operators $\alpha_{\bm{k}}$ and $\beta_{\bm{k}}$, we have
\begin{eqnarray}
\hat{n}_{\bm{k}}^{(a)}&=&\cosh^2(\theta_{\bm{k}})\alpha_{\bm{k}}^\dagger\alpha_{\bm{k}}\nonumber\\
&-&\sinh(\theta_{\bm{k}})\cosh(\theta_{\bm{k}})(\alpha_{\bm{k}}^\dagger\beta_{\bm{k}}^\dagger\nonumber\\
&+&\alpha_{\bm{k}}\beta_{\bm{k}})+\sinh^2(\theta_{\bm{k}})(\beta^\dagger_{\bm{k}}\beta_{\bm{k}}+1),\\
\hat{n}_{\bm{k}}^{(b)}&=&\sinh^2(\theta_{\bm{k}})(\alpha_{\bm{k}}^\dagger\alpha_{\bm{k}}+1)\nonumber\\
&-&\sinh(\theta_{\bm{k}})\cosh(\theta_{\bm{k}})(\alpha_{\bm{k}}^\dagger\beta_{\bm{k}}^\dagger+\alpha_{\bm{k}}\beta_{\bm{k}})\nonumber\\
&+&\cosh^2(\theta_{\bm{k}})\beta^\dagger_{\bm{k}}\beta_{\bm{k}}.
\end{eqnarray}
Particularly, if we start from the ground state, $\langle\alpha_{\bm{k}}^\dagger\alpha_{\bm{k}}(0)\rangle=\langle\beta_{\bm{k}}^\dagger\beta_{\bm{k}}(0)\rangle=\langle\alpha_{\bm{k}}\beta_{\bm{k}}(0)\rangle=0$, we find
\begin{eqnarray*}
\langle\hat{n}_{\bm{k}}^{(a)}(t)\rangle=\langle\hat{n}_{\bm{k}}^{(b)}(t)\rangle&=&\cosh(2\theta_{\bm{k}})\bar{n}_{\bm{k}}\left(1-{\rm e}^{-\gamma_{\bm{k}}t}\right)\\
&+&\sinh^2(\theta_{\bm{k}}).
\end{eqnarray*}
Introducing these values in (\ref{stm}) and using equation (\ref{xi}),
\begin{equation}\label{mag}
m^\mathrm{st}=m^\mathrm{st}_0-\frac{d}{8\pi^3}\int_\mathrm{B.Z.} d\bm{k}\left(\frac{\bar{n}_{\bm{k}}}{\sqrt{d^2-\xi_{\bm{k}}^2}}\right)\left(1-{\rm e}^{-\gamma_{\bm{k}}t}\right),
\end{equation}
where
\begin{equation}
m^\mathrm{st}_0=S-\frac{1}{16\pi^3}\int_\mathrm{B.Z.} d\bm{k}\left(\frac{d}{\sqrt{d^2-\xi_{\bm{k}}^2}}-1\right)
\end{equation}
is the expectation value of the staggered magnetization in the ground state. In 3D, for a square lattice and $S=1/2$, this value is $m^\mathrm{st}_0\simeq0.422$.

\begin{figure}[t]
\begin{center}
\includegraphics[width=0.48\textwidth]{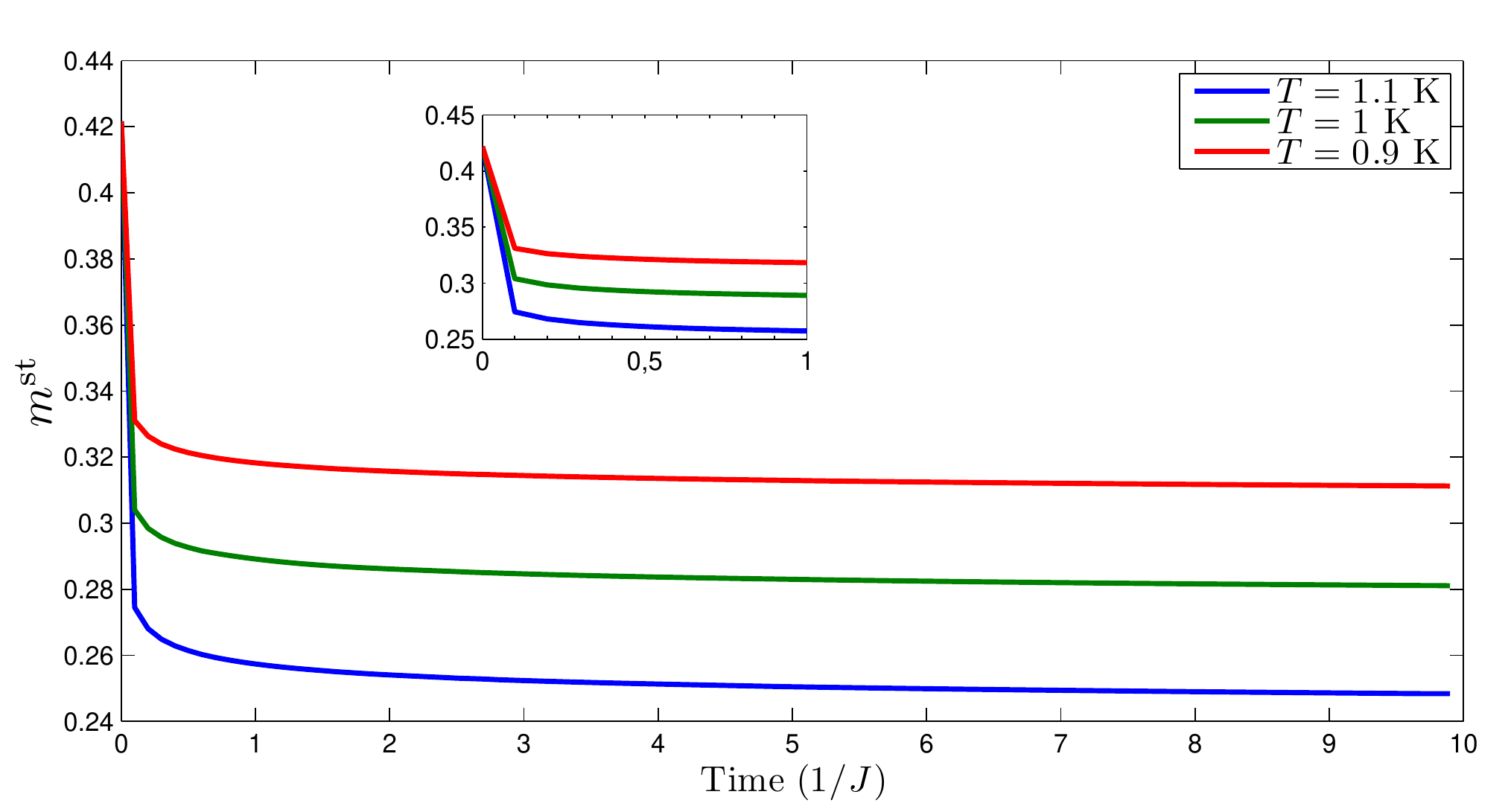}
\end{center}
\caption{Decay of the staggered magnetization from the ground state showing the approach to the Gibbs state values $m^\mathrm{st}_\beta$. The red line corresponds to $T=0.9$ K with $m^\mathrm{st}_\beta\simeq0.302$; for the green line, $T=1$ K and $m^\mathrm{st}_\beta\simeq0.271$; and for the blue line, $T=1.1$ K and $m^\mathrm{st}_\beta\simeq0.237$. The inset shows the evolution at short times in more detail.}\label{fig3}
\end{figure}

In figure \ref{fig3}, the evolution of the staggered magnetization is shown for different values of the bath temperature. It is noteworthy to mention the non-exponential decay of $m^{\mathrm{st}}$. This is due to its dependence on $t$ through the integral of (\ref{mag}), which renders combinations of different exponentials. Remarkably, there is a short period, where the order is lost very fast (between $t=0$ and $t\sim0.1/J$, see the inset figure). After that, the system continues evolving slower to the Gibbs state. This suggests that if we want to visualize variations of $m^{\rm st}$ due to the environment, the best chance is to look for them in systems with not very small $J$. Other thermal initial conditions lead to similar evolution in the magnetization.

\subsection{Two-point  correlation functions}

It is also worthwhile to study the second moments of angular momentum operators. For the sake of illustration, we focus in this section on the transversal two-point spatial correlation function, which is
\begin{equation}
S_\bot(\bm{r}_1,\bm{r}_2,t)=\mathrm{Tr}[S_{\bm{r}_1}^xS_{\bm{r}_2}^x\rho(t)].
\end{equation}
Without loss of generality, we take $\bm{r}_1\in A$. Then, for $\bm{r}_2\in A$,
\begin{equation}
S_\bot(\bm{r}_1,\bm{r}_2,t)=\frac{S}{2}\mathrm{Tr}[(a_{\bm{r}_1}+a_{\bm{r}_1}^\dagger)(a_{\bm{r}_2}+a_{\bm{r}_2}^\dagger)\rho(t)],
\end{equation}
and
\begin{equation}
S_\bot(\bm{r}_1,\bm{r}_2,t)=\frac{S}{2}\mathrm{Tr}[(a_{\bm{r}_1}+a_{\bm{r}_1}^\dagger)(b_{\bm{r}_2}+b_{\bm{r}_2}^\dagger)\rho(t)],
\end{equation}
for $\bm{r}_2\in B$.

\begin{figure}[t]
\begin{center}
\includegraphics[width=0.5\textwidth]{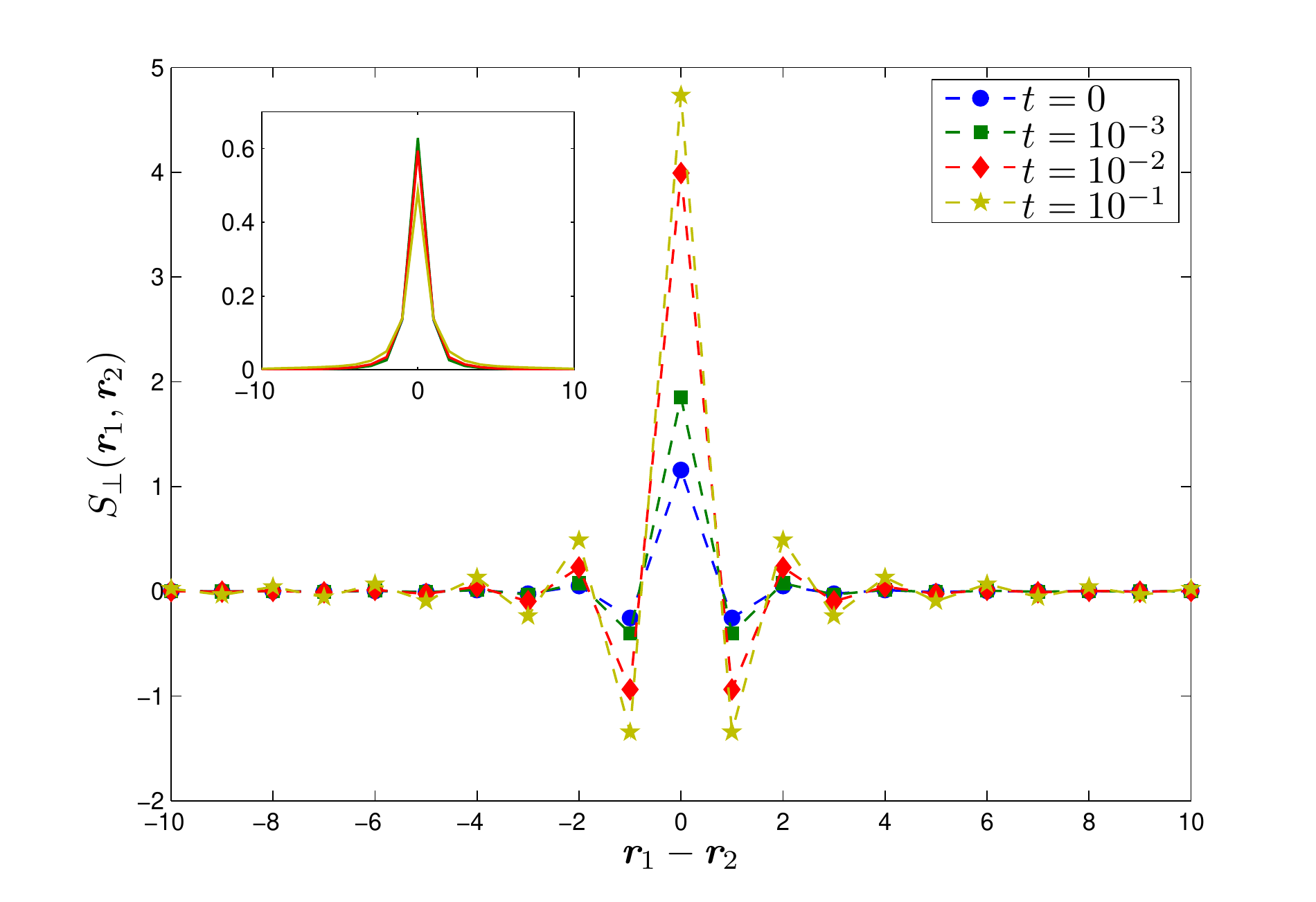}
\end{center}
\caption{Evolution of $S_\bot(\bm{r}_1,\bm{r}_2,t)$ from the ground state for a bath with $T=5$ K in the thermodynamic limit for different time instants (in units of $J^{-1}$). Note the oscillating behavior typical of antiferromagnetic systems. The inset figure illustrates the similarity between the cases for $|S_\bot(\bm{r}_1,\bm{r}_2,t)|$ after normalization.}\label{fig4}
\end{figure}

\begin{figure}[t]
\begin{center}
\includegraphics[width=0.5\textwidth]{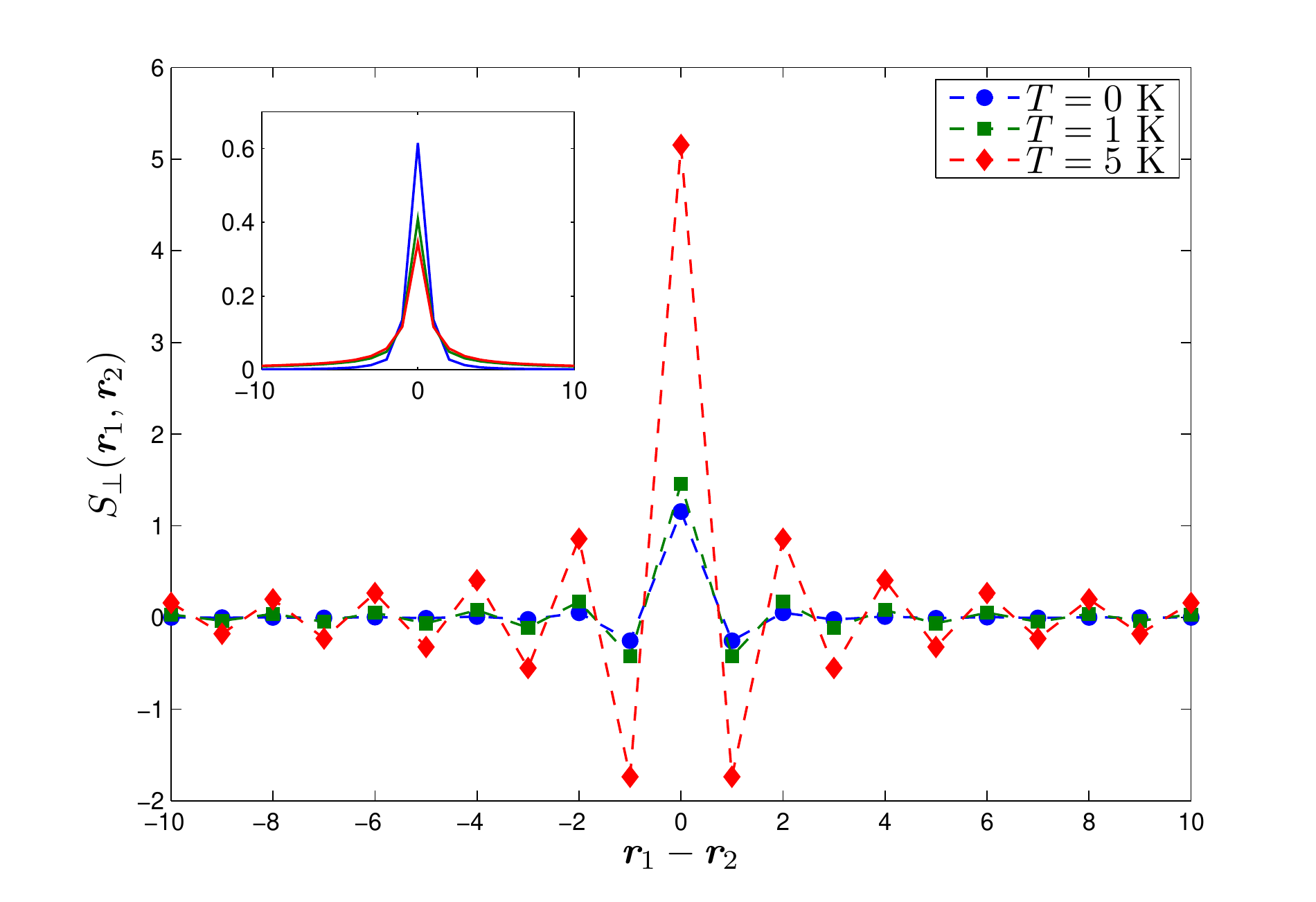}
\end{center}
\caption{Thermal values of $S_\bot(\bm{r}_1,\bm{r}_2)$. The inset represents again its normalized absolute value.}\label{fig5}
\end{figure}

Details of the computation are found in Appendix \ref{appendixB}; finally in the thermodynamic limit, we obtain
\begin{widetext}
\begin{equation}
S_\bot(\bm{r}_1,\bm{r}_2,t)=\frac{2S}{(2\pi)^3}\int_\mathrm{B.Z.} d\bm{k}\cos[\bm{k}\cdot(\bm{r}_1-\bm{r}_2)]\Theta_{\bm{k}}(\bm{r}_2)\left[\frac{2\bar{n}_{\bm{k}}\left(1-{\rm e}^{-\gamma_{\bm{k}}t}\right)+1}{\sqrt{d^2-\xi_{\bm{k}}^2 }}\right],
\end{equation}
\end{widetext}
where
\begin{equation}
\Theta_{\bm{k}}(\bm{r})=\begin{cases}d, &\text{if }\bm{r}\in A,\\
-\xi_{\bm{k}}, &\text{if }\bm{r}\in B.
\end{cases}
\end{equation}
We have plotted this correlation for some time instants in figure \ref{fig4}.
In addition, figure \ref{fig5} shows different cases when the Gibbs state has been reached.

\subsection{Response function}

Other interesting quantities in this system are the response functions. They are the Fourier transform of two-time correlation functions of spin operators, and for instance, they directly appear
in cross-sections of inelastic neutron scattering, which are experimentally accessible. For an antiferromagnet with staggered magnetization in the $z$ direction the inelastic scattering is related to the correlation  $\langle S^x_{-\bm{k}}(t+\tau)S^x_{\bm{k}}(t)\rangle$, where
\begin{equation}\label{sxk}
S^x_{\bm{k}}=\frac{1}{\sqrt{N}}\sum_{\bm{r}}{\rm e}^{{\rm i}\bm{k}\cdot\bm{r}}S_{\bm{r}}^x.
\end{equation}
One has to be especially careful when computing multitime-correlation functions for non-unitary evolutions. This is because the evolution of the product of two operators, say $a$ and $b$, is not equal to the product of the individual evolutions of $a$ and $b$ when the dynamics is not unitary, i.e. $(ab)(t)\neq a(t)b(t)$. However, we can circumvent this problem by writing the correlation function on the extended space where the evolution is indeed unitary:
\begin{eqnarray*}
\langle0| S^x_{-\bm{k}}(t+\tau)S^x_{\bm{k}}(t)|0\rangle=\mathrm{Tr}\langle[S^x_{-\bm{k}}(t+\tau)S^x_{\bm{k}}(t)|0\rangle\langle0|\otimes\rho_E]\nonumber\\
=\mathrm{Tr}\left[{\rm e}^{{\rm i}H(t+\tau)}S^x_{-\bm{k}}{\rm e}^{-{\rm i}H\tau}S^x_{\bm{k}}{\rm e}^{-{\rm i}Ht}|0\rangle\langle0|\otimes\rho_E\right].
\end{eqnarray*}
Here, the trace operation is taken over both the system and the environment degrees of freedom, and $H=H_{LSW}+H_E+V_{LSW}$ is the whole Hamiltonian of the system and the environment. Then, it is possible to obtain that (see the detailed discussion in \cite{GardinerZoller})
\begin{equation}
\langle0| S^x_{-\bm{k}}(t+\tau)S^x_{\bm{k}}(t)|0\rangle=\langle0|\left[S^x_{-\bm{k}}(\tau)S^x_{\bm{k}}\right](t)|0\rangle.
\end{equation}
That  is, it is needed to obtain first the Heisenberg evolution with respect to the parameter $\tau$ of the operator $S^x_{-\bm{k}}$ and after that the Heisenberg evolution with respect to the parameter $t$ of the product $S^x_{-\bm{k}}(\tau)S^x_{\bm{k}}$.

For linear spin-wave theory, we have
\begin{equation}
S_{\bm{r}}^x=\frac{S_{\bm{r}}^++S_{\bm{r}}^-}{2}=\sqrt{\frac{S}{2}}\begin{cases}
a_{\bm{r}}+a_{\bm{r}}^\dagger,&\text{if }\bm{r}\in A,\\
b_{\bm{r}}+b_{\bm{r}}^\dagger,&\text{if }\bm{r}\in B,
\end{cases}
\end{equation}
thus, according to (\ref{fourierab}) and (\ref{sxk}),
\begin{equation}
S^x_{\bm{k}}=\frac{\sqrt{S}}{2}(a_{\bm{k}}+a^\dagger_{-\bm{k}}+b_{-\bm{k}}+b^\dagger_{\bm{k}}).
\end{equation}
If we perform the Bogoliubov transformation (\ref{Bologiubova}) and (\ref{Bologiubovb}), the Eqs. (\ref{alpha(t)}) and (\ref{beta(t)}) lead to
\begin{widetext}
\begin{eqnarray}
S^x_{-\bm{k}}(\tau)&=&\frac{\sqrt{S}}{2}{\rm e}^{-\gamma_{\bm{k}}\tau/2}\left\{{\rm e}^{-{\rm i}\omega(\bm{k})\tau}\left[\cosh(\theta_{\bm{k}})\alpha_{-\bm{k}}-\sinh(\theta_{\bm{k}})\beta_{\bm{k}}+\cosh(\theta_{\bm{k}})\beta_{\bm{k}}-\sinh(\theta_{\bm{k}})\alpha_{-\bm{k}}\right]\right.\nonumber\\
&+&\left.{\rm e}^{{\rm i}\omega(\bm{k})\tau}\left[\cosh(\theta_{\bm{k}})\alpha^\dagger_{\bm{k}}+\cosh(\theta_{\bm{k}})\beta^\dagger_{-\bm{k}}-\sinh(\theta_{\bm{k}})\alpha^\dagger_{\bm{k}}-\sinh(\theta_{\bm{k}})\beta^\dagger_{-\bm{k}}\right]\right\},
\end{eqnarray}
where we have used the fact that $\theta_{-\bm{k}}=\theta_{\bm{k}}$ and $\omega(-\bm{k})=\omega(\bm{k})$. Since by assumption $\gamma_{\bm{k}}$ is small, for small $\tau$, we can neglect it in comparison to the complex exponential ${\rm e}^{\pm{\rm i}\omega(\bm{k})\tau}$:
\begin{equation}
S^x_{-\bm{k}}(\tau)\simeq\frac{\sqrt{S}}{2}[\cosh(\theta_{\bm{k}})-\sinh(\theta_{\bm{k}})]\left[{\rm e}^{-{\rm i}\omega(\bm{k})\tau}(\alpha_{-\bm{k}}+\beta_{\bm{k}})\right.
+\left.{\rm e}^{{\rm i}\omega(\bm{k})\tau}(\alpha^\dagger_{\bm{k}}+\beta^\dagger_{-\bm{k}})\right].
\end{equation}
Finally, by using (\ref{nalpha(t)}) and (\ref{nbeta(t)}), we compute the evolution of $S^x_{-\bm{k}}(\tau)S^x_{\bm{k}}$ with respect to $t$, and after simplifying vanishing terms, the correlation function reads
\begin{eqnarray}
\langle S^x_{-\bm{k}}(t+\tau)S^x_{\bm{k}}(t)\rangle&=&\frac{S}{4}[\cosh(2\theta_{\bm{k}})-\sinh(2\theta_{\bm{k}})]\left\{{\rm e}^{-{\rm i}\omega(\bm{k})\tau}[\langle\alpha_{-\bm{k}}\alpha^\dagger_{-\bm{k}}(t)\rangle+\langle\beta_{\bm{k}}\beta^\dagger_{\bm{k}}(t)\rangle]\right.\nonumber\\
&+&\left.{\rm e}^{{\rm i}\omega(\bm{k})\tau}[\langle\alpha^\dagger_{\bm{k}}\alpha_{\bm{k}}(t)\rangle+\langle\beta^\dagger_{-\bm{k}}\beta_{-\bm{k}}(t)\rangle]\right\}\nonumber\\
&=&\frac{S(d-\xi_{\bm{k}})}{2\sqrt{d^2-\xi^2_{\bm{k}}}}\left\{{\rm e}^{-{\rm i}\omega(\bm{k})\tau}\left[\bar{n}_{\bm{k}}\left(1-{\rm e}^{-\gamma_{\bm{k}}t}\right)+1\right]+{\rm e}^{{\rm i}\omega(\bm{k})\tau}\bar{n}_{\bm{k}}\left(1-{\rm e}^{-\gamma_{\bm{k}}t}\right)\right\}.
\end{eqnarray}
\end{widetext}
The Fourier transform with respect to $\tau$ leads to the response function
\begin{equation}
S_\bot(\bm{k},t,\omega)=S_{LSW}^{-}(\bm{k},t)\delta[\omega-\omega(\bm{k})]+S_{LSW}^{+}(\bm{k},t)\delta[\omega+\omega(\bm{k})],
\end{equation}
with
\begin{eqnarray}
S^{-}_{LSW}(\bm{k},t)&=&\frac{S(d-\xi_{\bm{k}})}{2\sqrt{d^2-\xi^2_{\bm{k}}}}\left[\bar{n}_{\bm{k}}\left(1-{\rm e}^{-\gamma_{\bm{k}}t}\right)+1\right],\\
S^{+}_{LSW}(\bm{k},t)&=&\frac{S(d-\xi_{\bm{k}})}{2\sqrt{d^2-\xi^2_{\bm{k}}}}\left[\bar{n}_{\bm{k}}\left(1-{\rm e}^{-\gamma_{\bm{k}}t}\right)\right].
\end{eqnarray}
Therefore, at $t=0$ (or $T=0$) ,only the form factor $S^{-}_{LSQ}(\bm{k},t)$ remains. On the other hand, we conclude that as temperature increases, $S^{\pm}_{LSQ}(\bm{k},t)$ also increases, and they have the same geometry in momentum space as the magnon decay rates $\gamma_{\bm{k}}$.

\section{Conclusions}
\label{sec:conclusions}

In this work, we have analyzed the behavior of a quantum antiferromagnet in contact with a boson thermal bath. Based on the spin-wave theory, we have applied the weak coupling procedure (Davies theory) to obtain a master equation for the dynamics. We believe that this is a basic and fundamental problem, which has remained quite unexplored so far.
It is at the crossroads of strongly correlated systems and the physics of open quantum systems that is so much rooted in quantum information theory.

From the open systems point of view, spin-wave theory provides us with a nice framework to apply the well-known techniques developed for quantum optics or quantum chemistry settings to quantum many-body problems. Interestingly, some features, which are typically encountered in small systems under weak coupling limit, e.g. the exponential decay of observables, may be lost when computing the observables, which are relevant for the many-body systems. We have exemplified this point by studying the staggered magnetization, which for moderate temperatures and despite of the isotropic coupling to the bath does not vanish. In fact, it does not show an exponential decay either.

Furthermore, we have illustrated the versatility of our master equation approach to the dynamics of thermal effects in quantum antiferromagnets by computing two-point correlation and response functions, also known as form factors. The geometry in momentum space of these response functions $S_{LSQ}(\bm{k},t)$ is closely related to that of the decay rate function in the first Brillouin zone.
These form factors, in turn, are directly related to differential cross-sections in experiments of inelastic neutron scattering, which, we believe, that may shed light to the current knowledge of a quantum antiferromagnet under non-isolated situations.

\begin{acknowledgments}
We thank the Spanish MICINN grant FIS2009-10061,
CAM research consortium QUITEMAD S2009-ESP-1594, European Commission
PICC: FP7 2007-2013, Grant No.~249958, UCM-BS grant GICC-910758.
\end{acknowledgments}

\appendix

\begin{widetext}

\section{Time-evolution of the first and second moments
}\label{appendixA}

The generator $\mathcal{L}^{\sharp}$ in the Heisenberg picture is obtained by the equality $\mathrm{Tr}[X\mathcal{L}(\rho)]=\mathrm{Tr}[\rho\mathcal{L}^\sharp(X)]$, for any operator $X$. By solving the dynamical equations, we obtain
\begin{eqnarray}
\alpha_{\bm{k}}(t)&=&{\rm e}^{[-{\rm i}\omega(\bm{k})-\gamma_{\bm{k}}/2]t}\alpha_{\bm{k}}(0),\label{alpha(t)}\\
\beta_{\bm{k}}(t)&=&{\rm e}^{[-{\rm i}\omega(\bm{k})-\gamma_{\bm{k}}/2]t}\beta_{\bm{k}}(0), \label{beta(t)} \\
\alpha_{\bm{k}}^\dagger\alpha_{\bm{k}}(t)&=&{\rm e}^{-\gamma_{\bm{k}}t}\alpha^\dagger_{\bm{k}}\alpha_{\bm{k}}(0)+
\bar{n}_{\bm{k}}[1-{\rm e}^{-\gamma_{\bm{k}}t}],\label{nalpha(t)}\\
\beta_{\bm{k}}^\dagger\beta_{\bm{k}}(t)&=&{\rm e}^{-\gamma_{\bm{k}}t}\beta^\dagger_{\bm{k}}\beta_{\bm{k}}(0)+
\bar{n}_{\bm{k}}[1-{\rm e}^{-\gamma_{\bm{k}}t}],\label{nbeta(t)},
\end{eqnarray}
The terms $\alpha_{\bm{k}}\alpha_{\bm{k}}^\dagger$ and $\beta_{\bm{k}}\beta_{\bm{k}}^\dagger$ are obtained by using the commutation relations $[\alpha_{\bm{k}},\alpha_{\bm{k}}^\dagger]=[\beta_{\bm{k}},\beta_{\bm{k}}^\dagger]=1$, and the remaining ones are just the composition of the dynamics given in \eqref{alpha(t)} and \eqref{beta(t)} and their Hermitian conjugate.

\section{Computation of the 2-spin correlator $S_\bot(\bm{r}_1,\bm{r}_2,t)$
}\label{appendixB}

We can compute the evolution of $S_\bot(\bm{r}_1,\bm{r}_2)$ in the Heisenberg picture. The terms $a_{\bm{r}_1}a_{\bm{r}_2}$, $a_{\bm{r}_1}b_{\bm{r}_2}^\dagger$ and their Hermitian conjugate do not contribute to the evolution. For $a_{\bm{r}_1}a^\dagger_{\bm{r}_2}$, we have
\begin{eqnarray}
(a_{\bm{r}_1}a_{\bm{r}_2}^\dagger)(t)&=&\frac{1}{N_A}\sum_{\bm{k},\bm{k}'}{\rm e}^{-{\rm i}\bm{k}\cdot\bm{r}_1}{\rm e}^{{\rm i}\bm{k}'\cdot\bm{r}_2}(a_{\bm{k}}a_{\bm{k}'}^\dagger)(t)\\
&=&\frac{1}{N_A}\sum_{\bm{k},\bm{k}'}{\rm e}^{-{\rm i}\bm{k}\cdot\bm{r}_1}{\rm e}^{{\rm i}\bm{k}'\cdot\bm{r}_2}\{[\cosh(\theta_{\bm{k}})\alpha_{\bm{k}}-\sinh(\theta_{\bm{k}})\beta^\dagger_{\bm{k}}][\cosh(\theta_{\bm{k}'})\alpha^\dagger_{\bm{k}'}-\sinh(\theta_{\bm{k}'})\beta_{\bm{k}'}]\}(t).\nonumber
\end{eqnarray}
By using Eqs. (\ref{alpha(t)})--(\ref{nbeta(t)}) and simplifying the mean values which vanish on the ground state, the sum is left only with the terms, where $\bm{k}=\bm{k}'$:
\begin{eqnarray}
\langle a_{\bm{r}_1}a_{\bm{r}_2}^\dagger(t)\rangle&=&\frac{1}{N_A}\sum_{\bm{k}}{\rm e}^{-{\rm i}\bm{k}\cdot(\bm{r}_1-\bm{r}_2)}[\cosh^2(\theta_{\bm{k}})\langle\alpha^\dagger_{\bm{k}}\alpha_{\bm{k}}(t)\rangle+\sinh^2(\theta_{\bm{k}})\langle\beta^\dagger_{\bm{k}}\beta_{\bm{k}}(t)\rangle+\cosh^2(\theta_{\bm{k}})]\nonumber\\
&=&\frac{1}{N_A}\sum_{\bm{k}}{\rm e}^{-{\rm i}\bm{k}\cdot(\bm{r}_1-\bm{r}_2)}[\cosh(2\theta_{\bm{k}})\bar{n}_{\bm{k}}\left(1-{\rm e}^{-\gamma_{\bm{k}}t}\right)+\cosh^2(\theta_{\bm{k}})].
\end{eqnarray}
Similarly, for the remaining non-vanishing terms, we obtain
\begin{eqnarray}
\langle a^\dagger_{\bm{r}_1}a_{\bm{r}_2}(t)\rangle&=&\frac{1}{N_A}\sum_{\bm{k}}{\rm e}^{-{\rm i}\bm{k}\cdot(\bm{r}_2-\bm{r}_1)}[\cosh^2(\theta_{\bm{k}})\langle\alpha^\dagger_{\bm{k}}\alpha_{\bm{k}}(t)\rangle+\sinh^2(\theta_{\bm{k}})\langle\beta^\dagger_{\bm{k}}\beta_{\bm{k}}(t)\rangle+\sinh^2(\theta_{\bm{k}})]\nonumber\\
&=&\frac{1}{N_A}\sum_{\bm{k}}{\rm e}^{-{\rm i}\bm{k}\cdot(\bm{r}_2-\bm{r}_1)}[\cosh(2\theta_{\bm{k}})\bar{n}_{\bm{k}}\left(1-{\rm e}^{-\gamma_{\bm{k}}t}\right)+\sinh^2(\theta_{\bm{k}})],
\end{eqnarray}
\begin{eqnarray}
\langle a_{\bm{r}_1}b_{\bm{r}_2}(t)\rangle&=&\langle a^\dagger_{\bm{r}_1}b^\dagger_{\bm{r}_2}(t)\rangle^\ast=\frac{-1}{N_A}\sum_{\bm{k}}{\rm e}^{-{\rm i}\bm{k}\cdot(\bm{r}_2-\bm{r}_1)}\cosh(\theta_{\bm{k}})\sinh(\theta_{\bm{k}})[\langle\alpha^\dagger_{\bm{k}}\alpha_{\bm{k}}(t)\rangle+\langle\beta^\dagger_{\bm{k}}\beta_{\bm{k}}(t)\rangle+1]\nonumber\\
&=&\frac{-1}{N_A}\sum_{\bm{k}}{\rm e}^{-{\rm i}\bm{k}\cdot(\bm{r}_2-\bm{r}_1)}\frac{\sinh(2\theta_{\bm{k}})}{2}[2\bar{n}_{\bm{k}}\left(1-{\rm e}^{-\gamma_{\bm{k}}t}\right)+1],
\end{eqnarray}
Thus, on the one hand, the correlation function for $\bm{r}_2\in A$ is
\begin{eqnarray}
S_\bot(\bm{r}_1,\bm{r}_2,t)&=&\frac{2S}{N_A}\sum_{\bm{k}}\cos[\bm{k}\cdot(\bm{r}_1-\bm{r}_2)]\cosh(2\theta_{\bm{k}})[2\bar{n}_{\bm{k}}\left(1-{\rm e}^{-\gamma_{\bm{k}}t}\right)+1]\nonumber\\
&+&\text{i}\frac{2S}{N_A}\sum_{\bm{k}}\sin[\bm{k}\cdot(\bm{r}_1-\bm{r}_2)].
\end{eqnarray}
Since
\begin{equation}
\frac{1}{N_A}\sum_{\bm{k}}\sin[\bm{k}\cdot(\bm{r}_1-\bm{r}_2)]=\mathrm{Im}\left\{\frac{1}{N_A}\sum_{\bm{k}}{\rm e}^{{\rm i}\bm{k}\cdot(\bm{r}_1-\bm{r}_2)}\right\}=\mathrm{Im}\left(\delta_{\bm{r}_1,\bm{r}_2}\right)=0,
\end{equation}
and using (\ref{xi}),
\begin{equation}
S_\bot(\bm{r}_1,\bm{r}_2,t)=\frac{2Sd}{N_A}\sum_{\bm{k}}\cos[\bm{k}\cdot(\bm{r}_1-\bm{r}_2)]\left[\frac{2\bar{n}_{\bm{k}}\left(1-{\rm e}^{-\gamma_{\bm{k}}t}\right)+1}{\sqrt{d^2-\xi_{\bm{k}}^2 }}\right].
\end{equation}
And on the other hand, for $\bm{r}_2\in B$,
\begin{eqnarray}
S_\bot(\bm{r}_1,\bm{r}_2,t)&=&\frac{-2S}{N_A}\sum_{\bm{k}}\cos[\bm{k}\cdot(\bm{r}_1-\bm{r}_2)]\sinh(2\theta_{\bm{k}})[2\bar{n}_{\bm{k}}\left(1-{\rm e}^{-\gamma_{\bm{k}}t}\right)+1]\nonumber\\
&=&\frac{-2S}{N_A}\sum_{\bm{k}}\cos[\bm{k}\cdot(\bm{r}_1-\bm{r}_2)]\frac{\xi_{\bm{k}}[2\bar{n}_{\bm{k}}\left(1-{\rm e}^{-\gamma_{\bm{k}}t}\right)+1]}{\sqrt{d^2-\xi_{\bm{k}}^2 }}.
\end{eqnarray}
\end{widetext}


\begin{thebibliography}{00}
\bibitem{RWA} By pressuposition we assume that the coupling is always small enough to safely neglect the counter-rotating terms for every $\bm{k}$, $\omega(\bm{k})t\gg g[\omega(\bm{k})]$. Note that there is not a problem with cases where $\omega(\bm{k})=0$, because the coupling, i.e. the spectral density, also approaches to zero, see Eqs. \eqref{decayrates} and \eqref{spectral_density}.
\end{thebibliography}
\end{document}